\documentclass[twocolumn,english,pra,floatfix,citeautoscript,nofootinbib,superscriptaddress]{revtex4-2}
\usepackage{dcolumn}
\usepackage{graphicx}
\usepackage{amsmath}
\usepackage{verbatim}
\usepackage{amssymb}
\usepackage{psfrag}
\usepackage{bm}
\usepackage{xcolor}
\usepackage{braket}
\usepackage{wrapfig}
\usepackage{subfigure}
\usepackage{makeidx}
\usepackage{bm}
\usepackage{epsf}
\usepackage{multirow}
\usepackage[colorlinks,linkcolor=blue,urlcolor=blue,citecolor=blue]{hyperref}

\begin{document}
\title{Collision Dynamics of Topologically Pumped Solitons}
\author{Ning Mao}
\affiliation{Laboratory of Quantum Information, University of Science and Technology of China, Hefei 230026, China}
\affiliation{Anhui Province Key Laboratory of Quantum Network, University of Science and Technology of China, Hefei 230026, China}
\affiliation{CAS Center For Excellence in Quantum Information and Quantum Physics, University of Science and Technology of China, Hefei 230026, China}
\author{Guang-Can Guo}
\affiliation{Laboratory of Quantum Information, University of Science and Technology of China, Hefei 230026, China}
\affiliation{Anhui Province Key Laboratory of Quantum Network, University of Science and Technology of China, Hefei 230026, China}
\affiliation{CAS Center For Excellence in Quantum Information and Quantum Physics, University of Science and Technology of China, Hefei 230026, China}
\affiliation{Hefei National Laboratory, University of Science and Technology of China, Hefei, 230088, China}
\author{Xi-Wang Luo}\email{luoxw@ustc.edu.cn}
\affiliation{Laboratory of Quantum Information, University of Science and Technology of China, Hefei 230026, China}
\affiliation{Anhui Province Key Laboratory of Quantum Network, University of Science and Technology of China, Hefei 230026, China}
\affiliation{CAS Center For Excellence in Quantum Information and Quantum Physics, University of Science and Technology of China, Hefei 230026, China}
\affiliation{Hefei National Laboratory, University of Science and Technology of China, Hefei, 230088, China}
\affiliation{Anhui Center for Fundamental Sciences in Theoretical Physics,
University of Science and Technology of China, Hefei 230026, China}
	
	

\begin{abstract}		
    Soliton pumping enables controlled transport of localized wave packets through the interplay between nonlinearity and band topology. Interactions between such wave packets provide an additional means to manipulate their localization and subsequent motion. Here, we investigate collisions of solitons in a driven bichromatic lattice, where solitons associated with bands of opposite Chern numbers are pumped toward one another. For intracomponent collisions, the soliton that dominates the collision survives as a localized outgoing wave packet, while the weaker one can lose its distinct localized identity. For intercomponent collisions, attractive interaction produces spatial locking, whereas repulsive interaction induces redistribution toward the lowest band and reverses the motion of the initially excited-band wave packet. Remarkably, for collisions between vector solitons with comparable initial intensities, the difference between their internal relative phases controls the collision outcome, favoring capture for in-phase preparation and re-separation for out-of-phase preparation. These results connect nonlinear scattering with subsequent topological pumping and identify initial intensity, intercomponent interaction, and relative phase as control parameters for manipulating localized transport within the same pumping protocol.
\end{abstract}
	
\maketitle
	
\section{Introduction}

Thouless pumping connects band topology to directed transport through cyclic adiabatic modulation \cite{Thouless1983,Niu1984,Xiao2010,Asboth2016,Citro2023,Lohse2016,Nakajima2016,Schweizer2016,Cerjan2020}. In the linear regime, quantized transport is typically formulated for uniformly occupied bands and does not necessarily preserve the localization of an individual wave packet \cite{Juergensen2021,Ke2016,Jurgensen2022Chern,Hu2019Dispersion,Fu2022Nonlinear}. Nonlinearity can counteract dispersion and thereby enable localized soliton transport while retaining the topological pumping response \cite{Juergensen2021,Fu2022Nonlinear}. This nonlinear setting has further revealed fractional pumping \cite{Jurgensen2023Fractional} and various forms of transport beyond linear-band predictions \cite{Jiang2026Intersite,Tao2026Anomalous,Tao2025Fractional}. The combination of spatial localization and directed motion makes solitons attractive for transporting and manipulating wave signals.

Controlled transport can also bring localized excitations into contact, where interactions provide a means of manipulating their quantum or wave states. In atomic quantum systems, Thouless pumping of individual atoms has been used to bring particles together for interaction-mediated state manipulation and quantum gates \cite{Zhu2025}, including protected geometric SWAP gates involving transient doublon occupation \cite{Kiefer2026}. For nonlinear waves, soliton collisions have been extensively studied in undriven or free-space systems, where integrable scalar solitons undergo elastic collisions while acquiring trajectory shifts~\cite{Zabusky1965,Gardner1967,Hirota1971,Zakharov1972}, and multicomponent solitons can exhibit phase-dependent trajectory changes, profile deformation, and redistribution among components \cite{Martin2007,Nguyen2014,Kanna2001ShapeChanging,Radhakrishnan1997,Vijayajayanthi2018}. However, the collision dynamics of solitons undergoing topological pumping remain largely unexplored. In such a driven setting, collisions occur while the wave packets are continuously transported by the lattice and can therefore reshape the localization, band composition, and pumping direction of the outgoing states.

In this work, we investigate the collision dynamics of scalar and vector solitons undergoing topological pumping in a driven bichromatic lattice. We prepare localized states associated with the lowest and second Bloch bands, which carry opposite Chern numbers and are pumped in opposite directions, bringing the two solitons toward one another. We first consider intracomponent collisions, where the relative initial intensities determine which localized wave packet survives the encounter and consequently select the direction of post-collision pumping. We then consider intercomponent collisions, where the sign of the intercomponent interaction provides an additional control knob: attractive interaction leads to spatial locking of the two solitons, whereas repulsive interaction redistributes the initially excited-band wave packet toward the lowest band and reverses its pumping direction. Finally, for vector solitons with comparable initial intensities, we show that the difference between their internal relative phases controls the collision outcome, allowing the outgoing wave packets to either merge into a dominant localized structure or re-separate. These results establish a direct connection between nonlinear scattering and subsequent topological pumping, showing that initial intensity, intercomponent interaction, and internal relative phase can be used to manipulate localized transport without changing the underlying pumping protocol.

\section{Model and methods}
The mean-field dynamics of a one-dimensional two-component Bose--Einstein condensate are governed by the coupled nonlinear Schr\"odinger equations \cite{Busch2001,Gubeskys2006,Lyu2024,Cao2025Vector},
\begin{eqnarray}\label{xde} i {\partial_t \psi_1} \!&=&\!\! \left[ -\frac{1}{2} {\partial^2_x} \!+\! V_{1}\left( x, t \right) \!+\! g_{11} |\psi_1|^2\! + \!g_{12} |\psi_2|^2 \right] \psi_1,\label{xde1} \\ i {\partial_t \psi_2} \!&\!\!\!=\!\!\!&\!\! \left[ -\frac{1}{2} {\partial^2_x}\!+\! V_{2}\left( x , t \right)\!+\! g_{12} |\psi_1|^2 \!+\! g_{22} |\psi_2|^2 \right] \psi_2.\label{xde2} \end{eqnarray}
Here, $\psi_j$ ($j=1,2$) denotes the condensate wavefunction of component $j$. The bichromatic lattice potential acting on each component is $V_{j}(x,t)
=-V_{sj}\cos^2\left(\frac{\pi x}{d}\right)
-V_{lj}\cos^2\left(\frac{\pi x}{2d}-\nu_j t\right)$ with $j=1,2$,
where $V_{sj}$ and $V_{lj}$ are the depths of the stationary and moving lattices, respectively, with $d$ and $2d$ their spatial periods, and $\nu_j$ the driving rate. 
Throughout this work, we focus on the case $V_1(x,t)=V_2(x,t)$ with $V_{s1}=V_{s2}\equiv V_{s}$, $V_{l1}=V_{l2}\equiv V_{l}$, and $\nu_1=\nu_2\equiv \nu$. 
The potential $V_j$ is periodic in both $x$ and $t$, with  lattice constant $a\equiv 2d$ and temporal period $T\equiv\pi/|\nu|$, respectively. The coefficients $g_{ij}$ characterize the intra- and intercomponent interactions, with positive (negative) values corresponding to repulsive (attractive) interactions.

In the linear limit, the Chern number of an isolated band determines the displacement of its Wannier center over one adiabatic pumping cycle \cite{Thouless1983}. A localized nonlinear state can inherit this displacement when it adiabatically follows a stable soliton branch connected to the corresponding linear band \cite{Juergensen2021,Mostaan2022}. This correspondence depends on the interaction strength and driving rate and is verified in the parameter regime considered in this work. Importantly, the Chern number constrains the net displacement over a pumping cycle rather than the detailed trajectory within the cycle \cite{Citro2023}. 
We use solitons associated with oppositely pumped bands and take their isolated evolution as a reference for identifying collision-induced changes in subsequent transport.

The initial solitons are obtained as stationary solutions of the nonlinear equations with the lattice frozen at the initial driving phase. For the lowest-band soliton, we employ imaginary-time evolution \cite{Bao2004,Yang2010}, starting from a hyperbolic-secant wave packet $\operatorname{sech}(x)$. For the second-band excited soliton, we use Newton iteration \cite{Yang2009,Yang2010} at a fixed chemical potential in the gap below the second band, starting from a second-band Wannier function $\mathcal{W}_2(x)$ multiplied by a hyperbolic-secant envelope $\operatorname{sech}[x/3]\mathcal{W}_2(x)$. After preparation, the solitons are placed at separated positions and evolved under the time-dependent lattice to study their approach, collision, and subsequent motion.

 \begin{figure}[t]
	\centering
	\includegraphics[width=88mm]{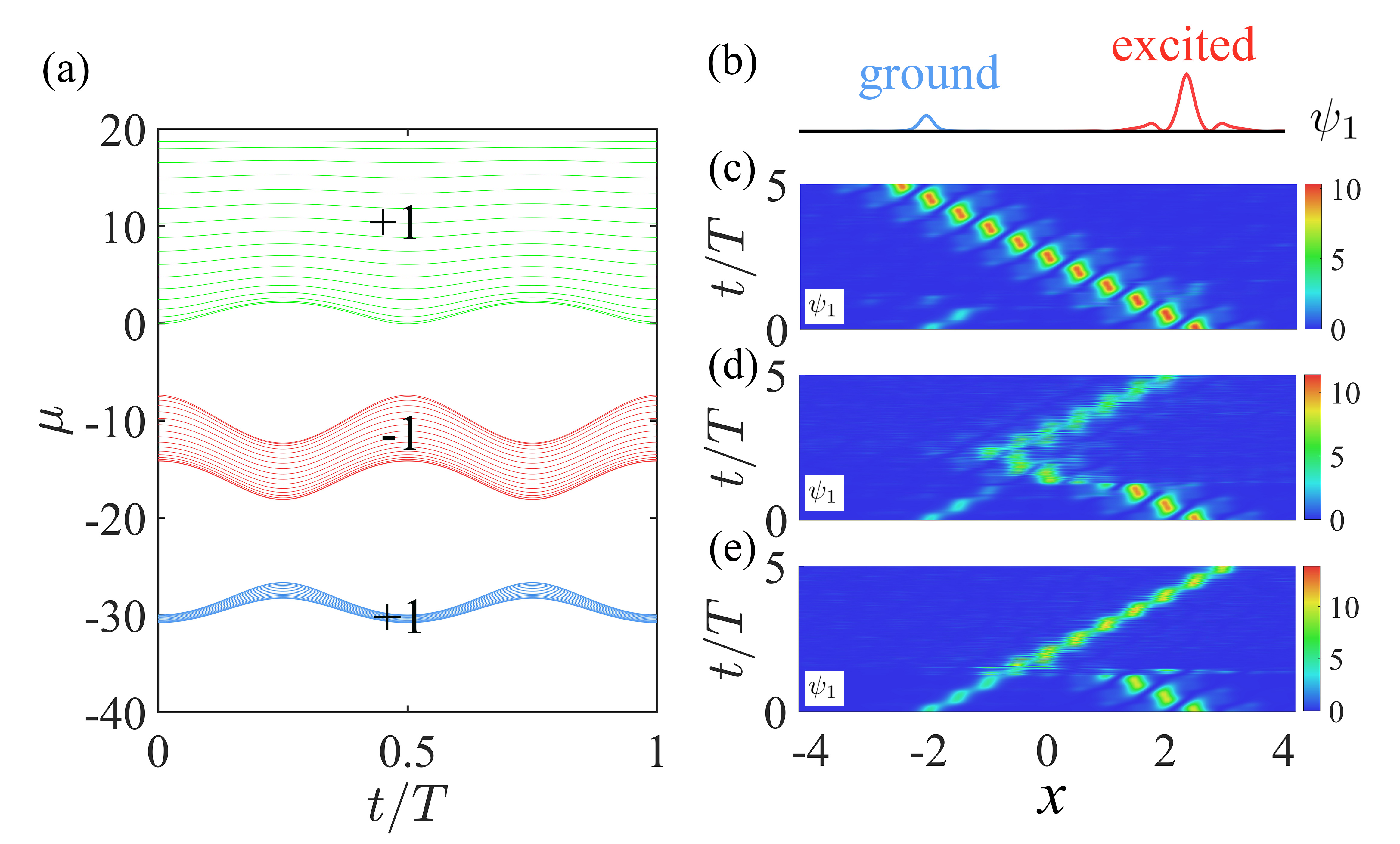}
\caption{Collisions of scalar solitons undergoing Thouless pumping.
	(a) Linear band spectrum over one pumping period, obtained
	from Eq.~\eqref{xde1} with $g_{11}=g_{12}=0$. The numbers
	indicate the Chern numbers of the corresponding bands.
	(b) Schematic profiles of the initially solitons $\tilde\psi_{g}(x)$ around $x=-2$ and $\tilde\psi_{e}(x)$ around $x=2.5$, associated with the lowest (blue) and second (red) bands, respectively.
	(c)-(e) Evolution of the density $|\psi_1(x,t)|^2$ with
    initial ground-soliton peak densities
	$\max_x|\tilde\psi_{g}(x)|^2=3$, $3.35$, and $5$, respectively. The excited-soliton peak density is $\max_x|\tilde\psi_{e}(x)|^2=10$, with corresponding chemical potential $\mu=-18$ in the first gap.
    The interaction is $g_{11}=-1$ in (b)-(e).
    Common parameters are: $V_{s}=25$, $V_{l}=25$,
	$d=1/2$, $\nu=0.05$, and $\psi_{2}=0$.
}\label{fig1}
\end{figure}

\section{Intracomponent collisions} 

We first consider a single-component system containing a ground-state soliton and an excited-state soliton. Specifically, we set $\psi_{2}=0$ and retain only the first component, reducing Eq.~\eqref{xde1} to a scalar nonlinear Schr\"odinger equation. The corresponding linear band structure, obtained by additionally setting $g_{11}=0$, is shown in Fig.~\ref{fig1}(a); details of the calculation can be found in Ref.~\cite{Cao2025Vector}. The lowest and second bands carry opposite Chern numbers defined in the parameter space $(k_x,t)$, with $C=+1$ and $C=-1$, respectively. In the regime where soliton pumping follows the Chern numbers of the associated linear bands, the two solitons are therefore pumped in opposite directions. As $t$ varies from $0$ to  $\pi/|\nu|$, corresponding to one pumping cycle, a soliton associated with a band of Chern number $C$ is displaced by $C$ unit cells along $x$. 

We initially place the two solitons at spatially separated positions such that they approach and collide during pumping, as illustrated in Fig.~\ref{fig1}(b). This configuration allows us to examine how a collision affects the subsequent localized transport. For conventional soliton collisions in integrable, undriven systems, the collisions preserve the asymptotic amplitudes and velocities of the incoming solitons while producing position and phase shifts \cite{Martin2007}. The topological pumping lattice considered here breaks this integrable structure, allowing collisions to generate radiation and alter the localization of the outgoing states. We therefore vary the initial intensity imbalance and relative phase of the incoming solitons to investigate their effects on the collision outcome and subsequent pumping. 

When the initial peak density of the ground-state soliton is relatively low, the excited-state soliton survives the collision and continues to move leftward, while the ground-state soliton loses its identity as a distinct localized excitation [Fig.~\ref{fig1}(c)]. As the peak density of the ground-state soliton increases, the collision produces pronounced deformation and dispersive radiation, leaving no clearly identifiable localized wave packet for the excited soliton after the collision [Fig.~\ref{fig1}(d)]. Interestingly, a portion of the particles may reform into a new localized wave packet in the ground bands, moving rightward during the post-collision pump. Upon a further increase in the ground-state peak density, the ground-state soliton dominates the collision, and its rightward-moving branch remains well localized, whereas the excited-state branch no longer persists as a distinct localized wave packet [Fig.~\ref{fig1}(e)].

Across the range of initial peak densities studied, the soliton that dominates the collision survives as a localized outgoing wave packet. This behavior contrasts with elastic collisions of integrable scalar solitons, in which both solitons retain their asymptotic amplitudes and velocities even when their initial intensities differ \cite{Zakharov1972}. In the driven lattice, the initial intensity imbalance therefore governs which localized branch survives and, consequently, the direction of its subsequent pumping. We characterize this transport by the displacement over one complete post-collision pumping cycle and present its dependence on the initial intensity imbalance in Appendix~\ref{secA1}. By contrast, varying the initial relative phase produces no appreciable change in the collision outcome or subsequent transport within the parameter range examined.

We next place the two incoming solitons in different components, introducing the intercomponent interaction as an additional control parameter with independently tunable strength and sign. In this configuration, the wave packets in the two components can overlap spatially without directly interfering, while still interacting through their densities. This allows us to examine how attractive and repulsive intercomponent interactions modify the collision outcomes and subsequent pumping.

\begin{figure}[t]
	\centering
	\includegraphics[width=88mm]{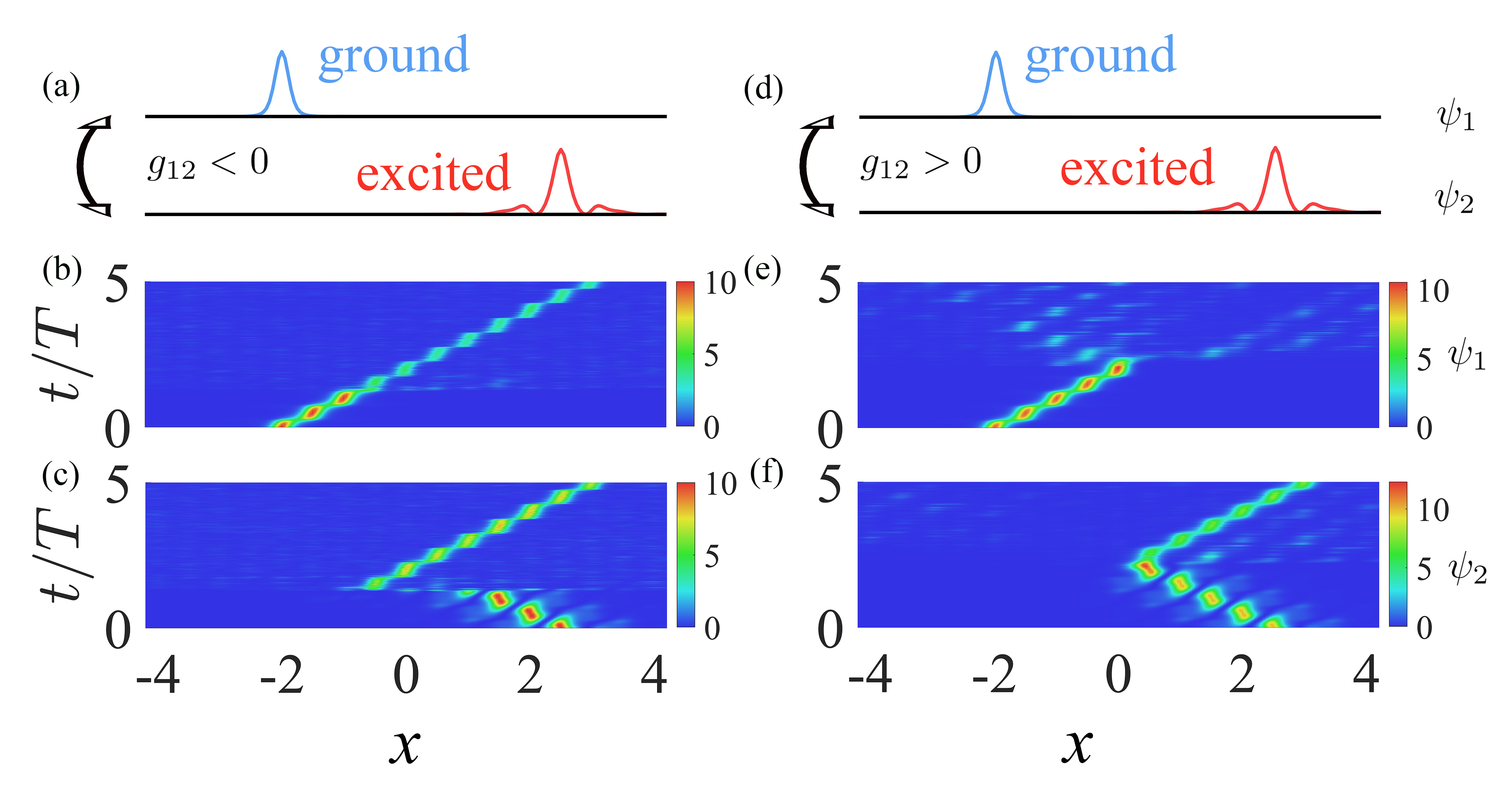}
\caption{Collisions of solitons in different components with
	(a)--(c) attractive interaction, $g_{12}=-5$, and
	(d)--(f) repulsive interaction, $g_{12}=5$.
	(a), (d) Schematic initial configurations, with the
	ground-state soliton in $\psi_1(x,0)=\tilde\psi_g(x)$ and the excited-state
	soliton in $\psi_2(x,0)=\tilde\psi_e(x)$.
	(b), (e) Density evolution of $|\psi_1|^2$;
	(c), (f) density evolution of $|\psi_2|^2$.
	We fix the intracomponent interaction strengths at $g_{11}=g_{22}=-1$ and the peak densities of the two solitons are $\max_x|\tilde\psi_g(x)|^2=\max_x|\tilde\psi_e(x)|^2=10$.
	The lattice parameters and driving rate are the same
	as in Fig.~\ref{fig1}.
}
	\label{fig2}
\end{figure}

\section{Intercomponent Collisions}

We consider collisions between two solitons that occupy different components of the two-component system. Specifically, a ground-state soliton is initialized in the first component and an excited-state soliton in the second component, at spatially separated positions [see Fig.~\ref{fig2}(a),(d)]. Since each incoming soliton occupies only one component, the two wave packets do not directly interfere. Instead, they interact through the intercomponent density-density interaction, which conserves the particle number of each component and makes the density dynamics invariant under a constant phase rotation of either component.

In the weak intercomponent-interaction limit, $|g_{12}|\to 0$, the solitons pass through each other with little change to their individual pumping dynamics. At stronger interaction and for markedly unequal initial intensities, the dominant soliton generally determines the subsequent localized transport, while the weaker soliton can undergo substantial deformation. Specifically, for repulsive interaction ($g_{12}>0$), the weaker soliton may lose its identity as a distinct localized excitation, whereas for attractive interaction ($g_{12}<0$), a small portion of the weaker soliton may be captured by the dominant one. Here, we focus on comparable initial intensities and investigate how attractive and repulsive intercomponent interactions lead to distinct collision outcomes.

\begin{figure}[t]
    \centering
    \includegraphics[width=90mm]{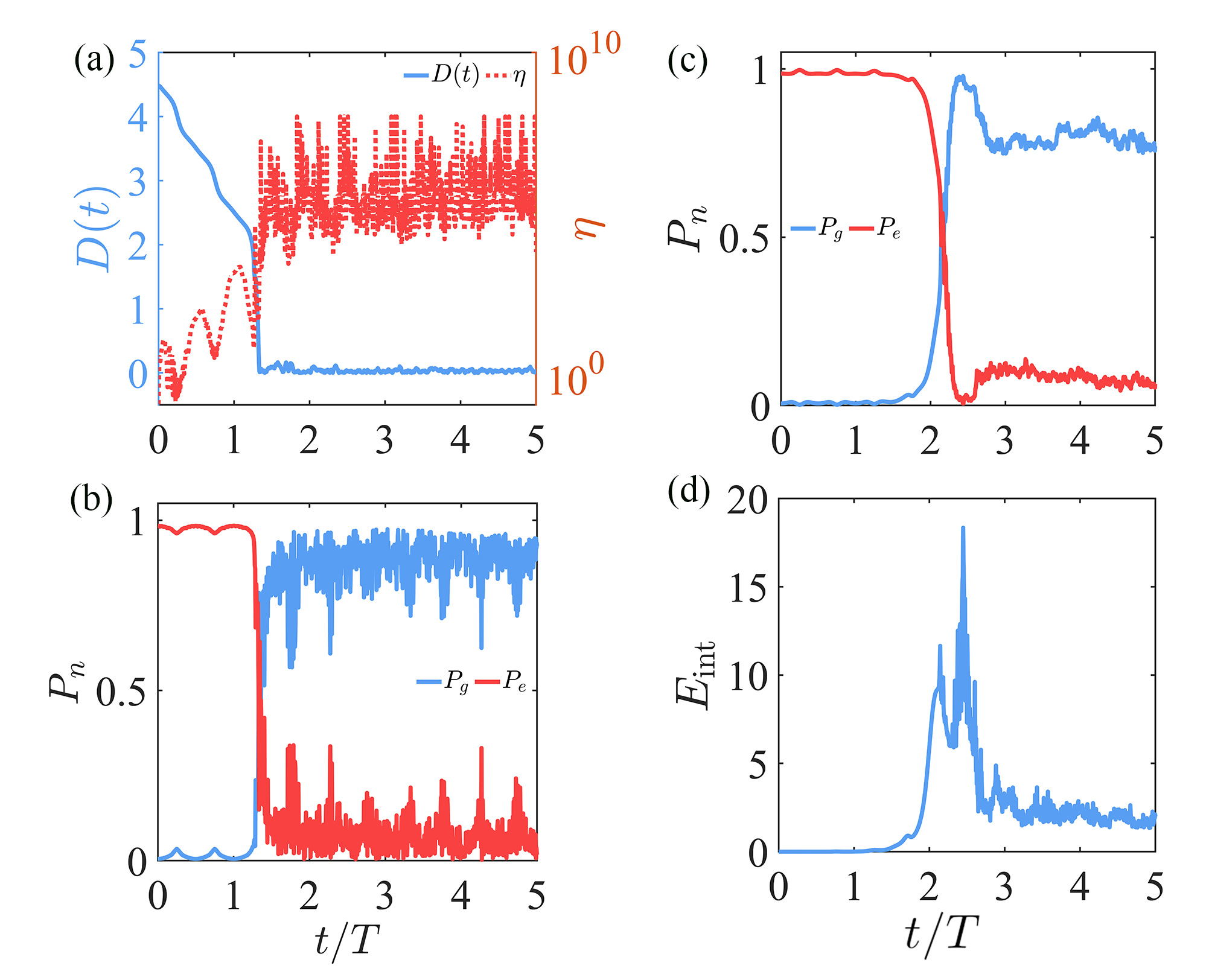}
    \caption{Collision diagnostics for attractive and repulsive intercomponent interaction.
(a) Core separation $D$ and energy ratio $\eta$ for attractive interaction. The small post-collision separation and large $\eta$ are consistent with spatial and velocity locking. {
(b), (c) Instantaneous first- and second-band weights ($P_g$ and $P_e$) of the second component wave packet $\psi_2$, for attractive and repulsive interaction, respectively.}
(d) Full-domain intercomponent interaction energy $E_{\mathrm{int}}$ for repulsive interaction. The rapid redistribution toward the first band in (c) occurs near the maximum of $E_{\mathrm{int}}$.
    }
    \label{fig3}
\end{figure}

(i) For attractive interaction, each soliton generates an attractive mean-field potential for the other. The density evolution in Figs.~\ref{fig2}(b),(c) shows that their encounter produces a spatially locked bright-bright composite in the ground band, with only a small fraction of the population transferred to dispersive radiation. The localized cores remain close and move together throughout the simulated post-collision pumping interval. To quantify this behavior, we track the two cores $X_j(t)$ (see Appendix~\ref{app:core_tracking} for the definition) in the coupled dynamics and evaluate their separation,
\begin{equation}
    D(t)=|X_2(t)-X_1(t)|.
\end{equation}
The persistently small separation in Fig.~\ref{fig3}(a), together with the joint localized motion, provides the principal evidence for capture. {As a complementary diagnostic, we consider the energy ratio
\begin{equation}
\eta(t)=\frac{E_{\mathrm{bind}}(t)}{K_{\mathrm{rel}}(t)},
\label{eq:capture_ratio}
\end{equation}
where $E_{\mathrm{bind}}$ characterizes the localized attractive intercomponent interaction energy and $K_{\mathrm{rel}}$ corresponds to the estimated kinetic energy of relative motion (see Appendix~\ref{app:core_energy} for more details). The large post-collision value of $\eta$ reflects suppressed relative motion and supports velocity locking (a regularization is applied for  vanishing $K_{\mathrm{rel}}$, see Appendix~\ref{app:core_energy}). }

Because the ground-state soliton is more stable than the excited-state soliton, attractive intercomponent interaction drives the latter toward the ground band during the collision, leading to the formation of a bright–bright composite. To characterize the band composition of the second component during the collision, we project $\psi_2(t)$ onto the instantaneous Bloch bands of the underlying linear lattice:
\begin{equation}
	P_n(t)=
	\frac{\langle\psi_2(t)|
		\hat{\mathcal P}_n(t)|\psi_2(t)\rangle}
	{\langle\psi_2(t)|\psi_2(t)\rangle},
\end{equation}
where $\hat{\mathcal P}_n(t)$ is the projection operator with $n=g$ and $n=e$ correspond to the projection onto the first (ground) and second (excited) band, respectively. The projection and normalization are evaluated
over the full spatial domain (thus $P_n$ includes contributions from both the localized wave packet and radiative waves).
The resulting weights, $P_g$ and $P_e$, quantify the contributions of the first and second bands to the second-component field, respectively. Before the collision, the second band dominates, with $P_e\simeq 1$ and $P_g\simeq 0$. During the collision, the population rapidly redistributes toward the first band, which remains dominant afterward [Fig.~\ref{fig3}(b)]. Together with the joint motion of the captured composite, this redistribution indicates a crossover from excited-band-dominated to ground-band-dominated transport in the second component. The resulting composite shares key characteristics with a vector soliton. Previous studies have shown that interactions can induce simultaneous pumping or arrest of two-component gap solitons \cite{Lyu2024}, while displaced spin-dependent lattices provide additional control over vector-soliton transport \cite{Cao2025Vector}. In our setting, the composite forms dynamically through the collision of initially separated, counterpropagating solitons, and its subsequent joint motion exhibits signatures of quantized pumping.

(ii) Repulsive intercomponent interaction ($g_{12}>0$) leads to a different outcome [Figs.~\ref{fig2}(d)--(f)]. After the encounter, the localized wave packet in $\psi_2$ reverses direction and moves toward increasing $x$, while the soliton in $\psi_1$ becomes strongly distorted and partially disperses. The accompanying band redistribution in $\psi_2$ is shown in Fig.~\ref{fig3}(c): $P_e$ rapidly decreases, while $P_g$ becomes dominant and remains near $0.8$ after the collision. This redistribution is consistent with the reversal of motion toward the pumping direction associated with the first band.

The band redistribution coincides with a pronounced increase in the intercomponent interaction energy,
\begin{equation}
    E_{\mathrm{int}}=g_{12}\int \rho_1(x,t)\rho_2(x,t)\,dx,
\end{equation}
during the encounter [Fig.~\ref{fig3}(d)], linking the redistribution to the collision-induced intercomponent interaction. The first component perturbs the second through the repulsive mean-field potential $g_{12}|\psi_1(x,t)|^2$, which evolves dynamically with the density of the first soliton. This mechanism differs from the scattering by a prescribed static impurity considered in Ref.~\cite{Cao2024ImpurityPumping}: here, the perturbing potential is generated by another evolving soliton, and both wave packets respond dynamically. The observed population transfer also connects to instability-mediated pump switching \cite{Jiang2026PumpSwitching}. However, the present results do not identify a specific Bogoliubov mode as the mechanism driving the transition.

In summary, attractive intercomponent interactions produce a jointly moving composite, whereas repulsive interactions reverse the motion of the initially excited-state wave packet through redistribution between the linear bands. Both outcomes are insensitive to a constant phase offset between the components, as expected for density–density interactions. We next turn to collisions between vector solitons, in which both incoming solitons occupy both components, allowing interference within each component to introduce a dependence on their relative phases.

\section{Vector Soliton Collisions}

We now consider collisions between ground-state and excited-state vector solitons, with both wave packets present in each component. Unlike the intercomponent collisions discussed above, the incoming vector solitons can interfere within each component during their encounter. This interference introduces a phase dependence into the collision dynamics and provides an additional means of controlling the outgoing localized structures.

\begin{figure}[t]
    \centering
    \includegraphics[width=88mm]{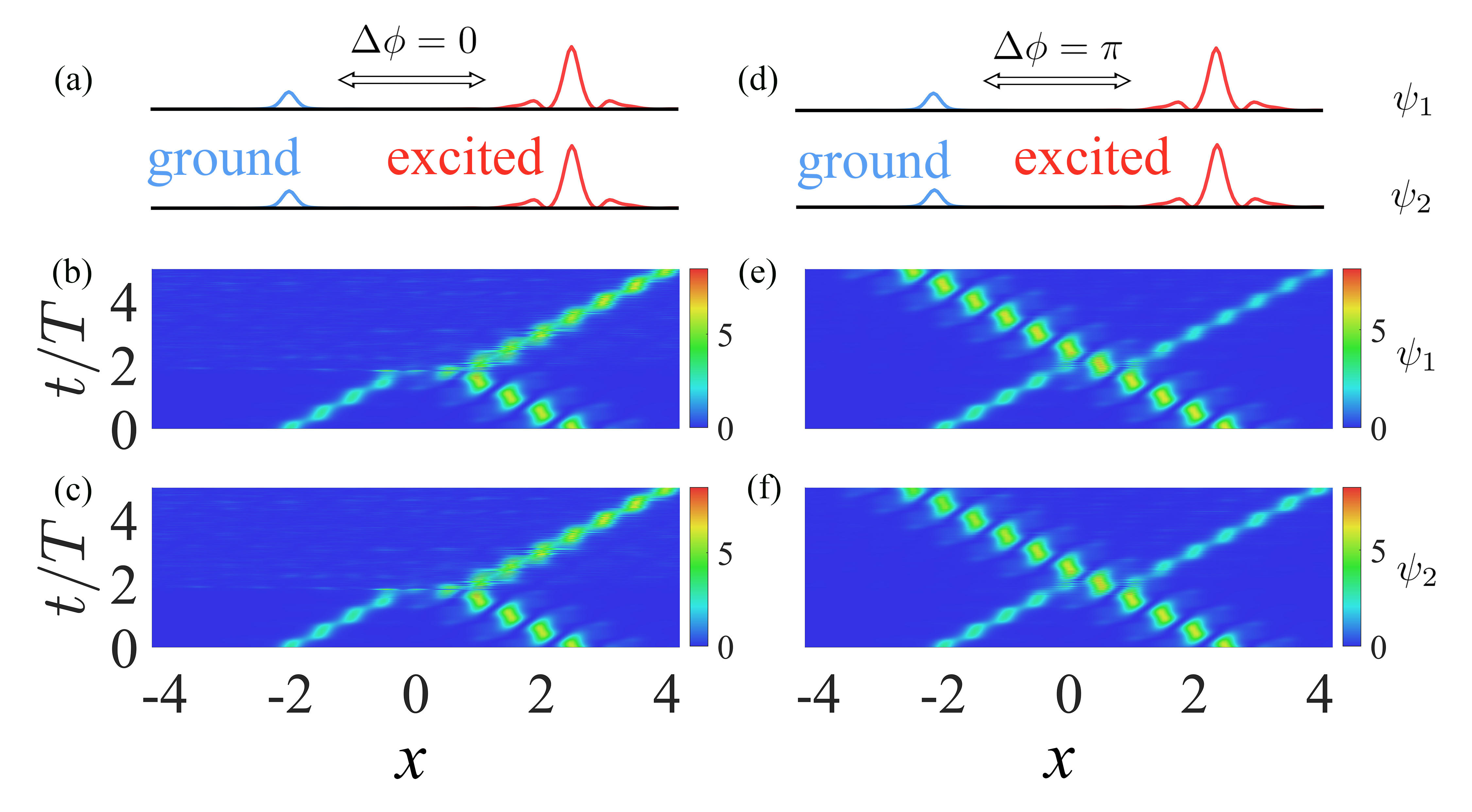}
    \caption{Phase-controlled collisions between ground- and excited-band vector solitons.
    (a), (d) Initial configurations for $\Delta\phi=0$ and $\Delta\phi=\pi$, respectively, with both wave packets present in each component.
    (b), (c) Density evolution of $|\psi_1|^2$ and $|\psi_2|^2$ for $\Delta\phi=0$, showing capture and a dominant outgoing localized wave packet.
    (e), (f) Corresponding evolution for $\Delta\phi=\pi$, showing survival and re-separation of the two localized wave packets.
    The ground- and excited-band solitons are initially centered at $x=-2$ and $x=2.5$, respectively, with equal populations in the two components.
    The interaction coefficients are $g_{11}=g_{22}=g_{12}=-1$.
    The initial peak densities are $\max_x|\tilde\psi_g(x)|^2=6$ for the ground-band soliton and $\max_x|\tilde\psi_e(x)|^2=12$ for the excited-band soliton.
    The lattice parameters and driving rate are the same as in Fig.~\ref{fig1}.
    }
    \label{fig4}
\end{figure}

\begin{figure}[t]
    \centering
    \includegraphics[width=88mm]{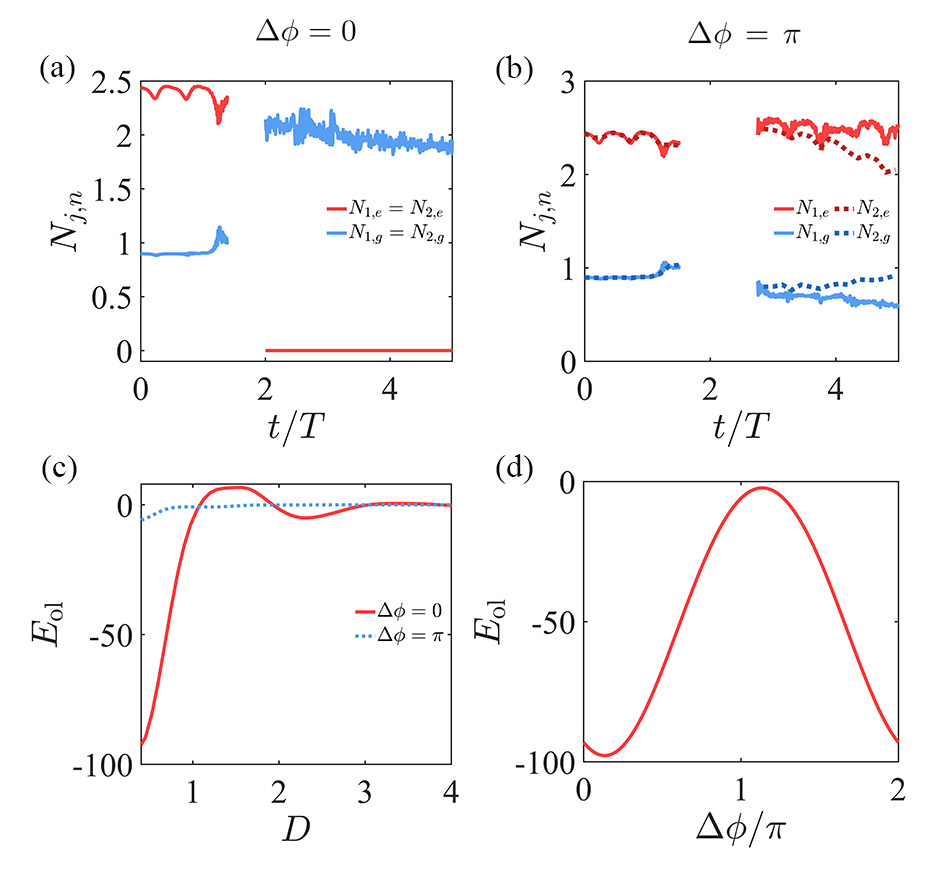}
    \caption{Localized populations and effective interaction energy in vector-soliton collisions.
    (a), (b) Localized populations $N_{j,n}$, corresponding to particle numbers within the tracked spatial windows, for $\Delta\phi=0$ and $\pi$, respectively. Blue and red denote the ground- and excited-state trajectories, while solid and dashed lines denote components $j=1$ and $2$. The localized populations are not well defined during the collision interval. For $\Delta\phi=0$, the surviving packet is tracked along the ground-state trajectory, with $N_{1,e}$ and $N_{2,e}$ approaching zero after the collision. For $\Delta\phi=\pi$, both outgoing packets retain nonzero localized populations.
    (c) Effective overlap energy $E_{\mathrm{ol}}$ versus separation $D$ for $\Delta\phi=0$ (red solid line) and $\pi$ (blue dotted line).
    (d) Dependence of $E_{\mathrm{ol}}$ on $\Delta\phi$ at the separation close enough ($D=0.4$). The overlap energy is more negative around $\Delta\phi=0$ than that around $\Delta\phi=\pi$.
    }
    \label{fig5}
\end{figure}

Relative-phase control of soliton interactions is well established in undriven systems, including optical fibers and matter-wave systems \cite{Gordon1983,Nguyen2014}. 
In the present topological pumping system, the density-density interaction between the components and the global U(1) gauge freedom ensure that a uniform phase rotation of an entire component, $\psi_j\to e^{i\theta_j}\psi_j$, leaves the dynamics unchanged. We can therefore fix the overall phases of the components when preparing the initial states. We use the excited-state vector soliton as the reference and consider a symmetric vector soliton of the form
\begin{equation}
    {\tilde\Psi}_e(x)=\tilde\psi_{e}(x) \begin{pmatrix}
       \frac{1}{\sqrt{2}} \\
       \frac{1}{\sqrt{2}} 
    \end{pmatrix},
\end{equation}
which exists over a broad range of interaction parameters \cite{Mao2022,Liu2009Formation}. Here, $\tilde{\psi}_e(x)$ denotes the spatial profile of the initially prepared excited-state soliton.

Similar to the scalar-soliton collisions, we find that varying the overall relative phase between the two incoming vector solitons does not appreciably affect the collision dynamics within the parameter range examined. We therefore fix the internal relative phase of the excited-state vector soliton and use the internal relative phase of the ground-state vector soliton as the control parameter. Specifically, we prepare the initial state of the ground-state vector soliton as
\begin{equation}
    \tilde{\Psi}_g(x)=\tilde{\psi}_g(x)
    \begin{pmatrix}
        \frac{e^{i\Delta\phi}}{\sqrt{2}}\\
        \frac{1}{\sqrt{2}}
    \end{pmatrix},
\end{equation}
where $\tilde{\psi}_g(x)$ is the spatial profile of the initially prepared ground-state soliton. Thus, $\Delta\phi$ specifies the internal relative phase of the ground-state vector soliton with respect to that of the excited-state vector soliton, whose internal relative phase is fixed to zero. Equivalently, the phase shift is applied only to the ground-state contribution in the first component.

The density in the each component contains a contribution from the interference between the two solitons,
\begin{equation}
    \begin{aligned}
        |\psi_j(x,t)|^2    =|\psi_{g,j}|^2+|\psi_{e,j}|^2+2\operatorname{Re}\!\left[
        \psi_{g,j}\psi_{e,j}^{*}
        \right].
    \end{aligned}
\end{equation}
Though the relative phase between the two solitons in the second component is fixed to zero in our preparation, a nonzero relative phase $\Delta\phi$ changes the interference term in the first component and may significantly modify the collision dynamics. Through the intercomponent density-density interaction, the interference affects the dynamics of both components. Although the wave packets are initially well separated and their interference is negligible, this phase dependence becomes important as they overlap during the collision. The sensitivity to $\Delta\phi$ depends on the initial intensity imbalance between the two solitons. When one soliton is much less intense than the other, the dominant outgoing localized structure generally follows the pumping direction of the dominant incoming soliton. In this regime, phase variations can modify the deformation and radiative losses but have limited influence on the qualitative transport outcome. At comparable initial intensities, however, the phase can determine whether the two wave packets are captured or re-separate. We therefore focus on this regime and compare $\Delta\phi=0$ and $\Delta\phi=\pi$ [see Figs.~\ref{fig4}(a),(d)].

To characterize the collision outcomes, we examine the
outgoing trajectories of the surviving solitons, which can
be seen directly in the density evolution. We also introduce the localized populations $N_{j,g}$ and $N_{j,e}$, defined as the particle numbers of component $j$ within the moving spatial windows associated with the ground- and excited-state trajectories. These populations track the localized parts of the wave packets rather than their Bloch-band occupations; their definitions and numerical implementation are given in Appendix~\ref{app:coherent_diagnostics}.
For $\Delta\phi=0$, the collision produces a dominant localized structure moving along the ground-state pumping direction, accompanied by weaker radiative waves [Figs.~\ref{fig4}(b),(c)]. After the encounter, $N_{1,e}$ and $N_{2,e}$ approach zero, while $N_{1,g}$ and $N_{2,g}$ increase [Fig.~\ref{fig5}(a)]. This redistribution indicates that the localized population becomes concentrated along the surviving ground-state trajectory, consistent with capture accompanied by reshaping and radiation.
For $\Delta\phi=\pi$, two localized wave packets survive and re-separate along approximately X-shaped trajectories [Figs.~\ref{fig4}(e),(f)]. Both outgoing packets retain nonzero localized populations in each component, although these populations differ from their initial values [Fig.~\ref{fig5}(b)], indicating inelastic redistribution during the encounter.

To provide an energetic perspective on the phase dependence, we evaluate the overlap energy $E_{\mathrm{ol}}(D,\Delta\phi)$ of the two vector solitons independently pumped to the collision time, with $D$ the manually imposed separation. Specifically, we evolve each vector soliton separately under the pumping protocol to the collision time $t_c$, determined by the coincidence of their centers. Denoting the resulting instantaneous profiles by
\begin{equation}
    \tilde{\Psi}_e^c(x)=
    \begin{pmatrix}
        \frac{\tilde{\psi}_e^c(x)}{\sqrt{2}}\\
        \frac{\tilde{\psi}_e^c(x)}{\sqrt{2}}
    \end{pmatrix},
    \qquad
    \tilde{\Psi}_g^c(x)=
    \begin{pmatrix}
        \frac{e^{i\Delta\phi}\tilde{\psi}_g^c(x)}{\sqrt{2}}\\
        \frac{\tilde{\psi}_g^c(x)}{\sqrt{2}}
    \end{pmatrix},
\end{equation}
we define the overlap energy as
$E_{\mathrm{ol}}(D,\Delta\phi)
=E[\tilde{\Psi}_e^c(x)+\tilde{\Psi}_g^c(x-D)]-E[\tilde{\Psi}_e^c(x)]
-E[\tilde{\Psi}_g^c(x-D)]$,
where $E[\Psi]$ is the mean-field energy functional of the coupled system. Further details are provided in Appendix~\ref{app:coherent_diagnostics}.
For $\Delta\phi=0$, $E_{\mathrm{ol}}$ develops a deep negative minimum at small separation [Fig.~\ref{fig5}(c)], indicating that overlap is energetically favorable for the profiles used in this diagnostic. For $\Delta\phi=\pi$, the overlap-energy curve is substantially shallower, consistent with a weaker tendency toward binding. At the selected separation, the overlap energy varies smoothly with $\Delta\phi$ and remains negative throughout the scan, with its minimum and maximum occurring at phases slightly larger than $0$ and $\pi$, respectively [Fig.~\ref{fig5}(d)]. This static energy comparison supports the observed phase dependence, while capture and re-separation are identified from the subsequent dynamical evolution of the localized structures.

Collisions of exact bright vector solitons can redistribute component intensities through polarization-dependent scattering coefficients \cite{Kanna2001ShapeChanging}, providing a basis for Manakov-soliton logic proposals \cite{Vijayajayanthi2018}. Here, the internal relative phase of the incoming ground-state vector soliton controls its interference with the excited-state wave packet in the first component. At comparable initial intensities, this phase control changes the number and trajectories of the outgoing localized structures, complementing the interaction-controlled locking and transport reversal found in intercomponent collisions. These phase-dependent outcomes suggest a route to routing localized waves under a fixed pumping protocol, connecting soliton-based logic proposals \cite{Vijayajayanthi2018} with controlled encounters in atomic transport circuits \cite{Zhu2025,Kiefer2026}. Implementing a logic gate would, however, require explicit encoding and a quantitative assessment of readout fidelity, radiative losses, and robustness.

\section{Conclusion and Discussion}

We have investigated collision-controlled transport of scalar and vector solitons undergoing topological pumping in a driven bichromatic lattice. By considering different types of collisions, we identify various mechanisms for controlling the localization and subsequent transport of the outgoing wave packets. For intracomponent collisions, the initial intensity imbalance determines which localized branch survives the encounter and thus selects the direction of subsequent pumping. For intercomponent collisions, attractive interactions can produce a spatially locked composite, whereas repulsive interactions redistribute the population toward the lowest Bloch band and reverse the motion of the initially excited-band wave packet. For vector-soliton collisions with comparable initial intensities, the difference between the internal relative phases of the two incoming vector solitons controls their collision outcome, favoring capture for in-phase $\Delta\phi=0$ preparation and re-separation for out-of-phase $\Delta\phi=\pi$ preparation.

These results connect topological pumping with nonlinear soliton scattering. In the weakly nonlinear regime, isolated solitons can exhibit quantized transport governed by the Chern number of their parent band \cite{Jurgensen2022Chern}, whereas stronger nonlinear interband interaction can arrest transport \cite{Fu2022Nonlinear}. 
Unlike in undriven integrable systems, where collisions between bright vector solitons can redistribute component intensities while preserving the total norm of each asymptotic soliton~\cite{Kanna2001ShapeChanging}, collisions in the driven topological pumping setting studied here provide additional control over subsequent transport by modifying the localization, band composition, and binding of the outgoing wave packets, without changing the underlying pumping protocol.

More broadly, our findings demonstrate how topological pumping and nonlinear scattering can be combined to manipulate localized excitations. Pumping brings solitons into contact, while collisions reorganize the states that carry subsequent robust transport. The resulting interplay offers a route to controlling localized-wave propagation through initial intensity, intercomponent interaction, and relative-phase engineering. Together with soliton-based logic proposals \cite{Vijayajayanthi2018} and experimental realizations of nonlinear Thouless pumping \cite{Juergensen2021}, these results motivate further investigation of programmable transport in interacting nonlinear wave systems. Extending this framework to multiple solitons, different lattice models, and spin-dependent topological pumping schemes could reveal how collective scattering, nonlinear binding, and band topology jointly govern transport, opening possibilities for controlled information flow and dynamical assembly in nonlinear wave systems.

\section*{Acknowledgments}
This work is supported by the National Natural Science Foundation of China (Grants No. 12574544), Quantum Science and Technology-National Science and Technology Major Project (Grant No. 2021ZD0301200), the CAS Project for Young Scientists in Basic Research (YSBR-085), and USTC start-up funding.
	
\section*{Data availability}
The data that support the findings of this article are publicly available~\cite{data}.	

	\begin{widetext}
	\begin{appendix}
		
		\counterwithout{equation}{section}
		\counterwithout{figure}{section}
		
		\setcounter{figure}{0} \renewcommand{\thefigure}{A\arabic{figure}} %
		\renewcommand*{\theHfigure}{appendix.\thefigure}

\section{Effects of relative intensity on intracomponent collision} 
\label{secA1}
\setcounter{equation}{0} \renewcommand{\theequation}{A\arabic{equation}}

To examine how the relative initial intensities affect transport after a scalar-soliton collision, we vary the initial peak density of the ground-state wave packet, \(\max_x|\tilde\psi_{g}(x)|^2\), while keeping the excited-state preparation and pumping protocol fixed. As a reference, Fig.~\ref{figA1}(a) shows the center trajectories of the two solitons evolved independently, without a collision. The ground- and excited-state solitons originate from Bloch bands with Chern numbers \(C_g=+1\) and \(C_e=-1\), respectively, and are pumped in opposite directions. For the lattice parameters \(d=1/2\), the common lattice constant is \(a=2d=1\). Over three pumping cycles, their displacements are approximately \(+3a=+3\) and \(-3a=-3\), corresponding to transport by one unit cell per cycle to the right and left, respectively.

For the collision dynamics, we track each surviving localized wave packet and evaluate its displacement over one complete pumping cycle,
\begin{equation}
	Q_{\rm out}(t_0)=\frac{x_{\rm out}(t_0+T)-x_{\rm out}(t_0)}{a},
\end{equation}
where \(x_{\rm out}(t)\) is the center position of the outgoing localized wave packet, and \(T=\pi/|\nu|\) is the pumping period. The reference time \(t_0\) is chosen after the collision, once the outgoing wave packet can be identified separately. Thus, \(Q_{\rm out}\simeq\pm1\) indicates that the surviving wave packet retains a displacement of approximately one unit cell per cycle in the corresponding direction.

\begin{figure}[htbp]
	\centering
	\includegraphics[width=88mm]{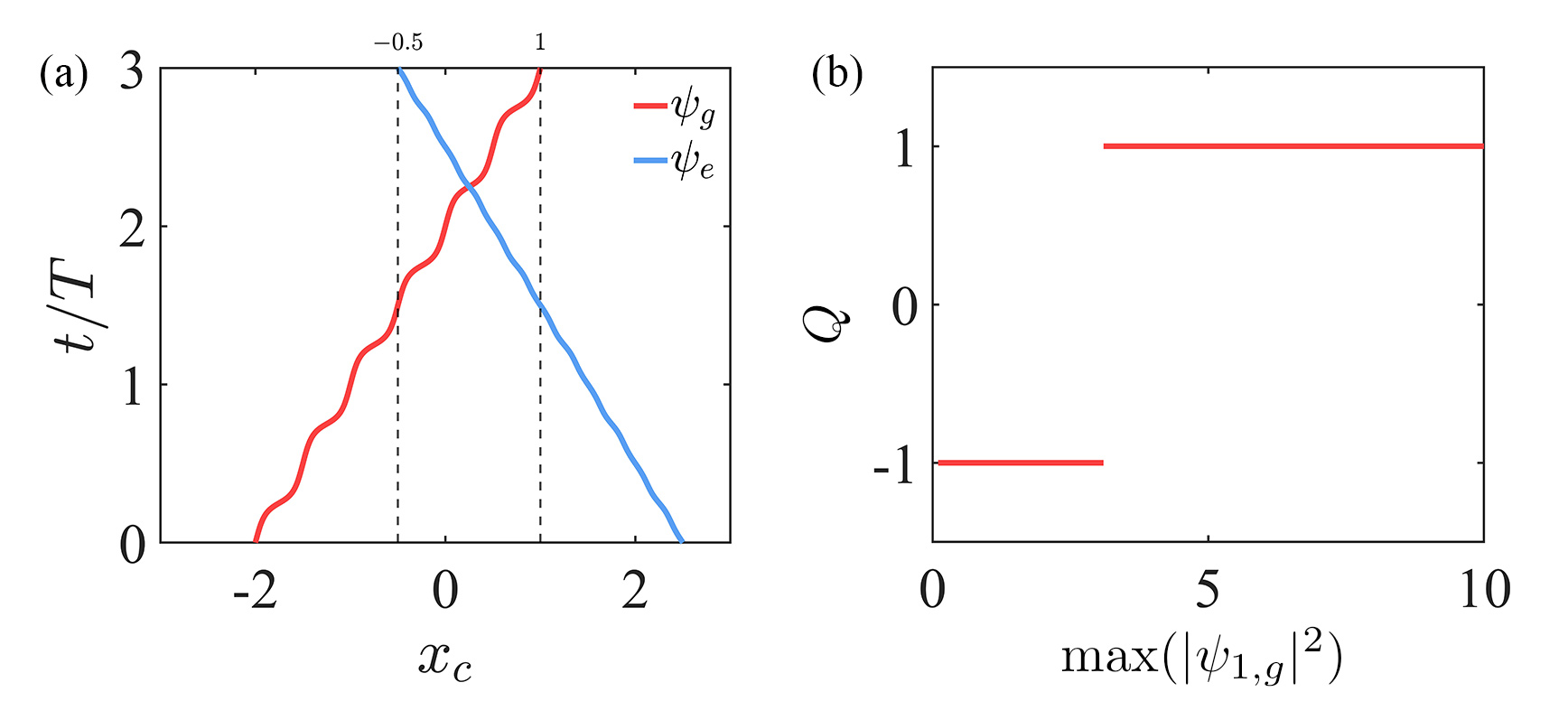}
	\caption{
		Initial-intensity dependence of post-collision pumping.
		(a) Center trajectories of the ground-state (red) and excited-state (blue) wave packets evolved independently, without a collision.
		(b) Displacement \(Q_{\rm out}\) of the surviving localized wave packet over one post-collision pumping cycle as a function of the initial ground-state peak density \(\max_x|\psi_{1,g}(x,0)|^2\). The excited-state preparation and pumping protocol are fixed. The plateaus at \(Q_{\rm out}\simeq-1\) and \(Q_{\rm out}\simeq+1\) correspond to transport by one lattice constant to the left and right, respectively.
	}
	\label{figA1}
\end{figure}

Figure.~\ref{figA1}(b) shows two plateaus in the displacement as the initial ground-state peak density is varied. At lower peak densities, the surviving wave packet moves to the left with \(Q_{\rm out}\simeq-1\), following the pumping direction of the incoming excited-state branch. Above a peak density of approximately \(3\), the outgoing localized wave packet instead moves to the right with \(Q_{\rm out}\simeq+1\), following the ground-state pumping direction. Thus, changing the relative initial intensities selects the direction of post-collision transport.

\section{Numerical diagnostics for inter-component and vector-soliton collisions}
\label{secA2}

\setcounter{equation}{0} \renewcommand{\theequation}{B\arabic{equation}}

This appendix collects the trajectory, population, and energy diagnostics used in Figs.~\ref{fig3} and \ref{fig5}. The core tracking and energy windows below apply to the intercomponent collision analysis. The spatial assignment used for vector soliton collisions is described separately in Appendix~\ref{app:coherent_diagnostics}.

\subsection{Adaptive core tracking}
\label{app:core_tracking}

For intercomponent collisions, to quantify spatial locking while reducing contributions from dispersive radiation, we identify the localized soliton core in each component directly from its instantaneous density profile, $\rho_j(x,t)=|\psi_j(x,t)|^2$. At each sampled time, an integration window $\mathcal{I}_j(t)$ is placed around the identified core, and its position is calculated as the local centroid,
\begin{equation}
	\widehat{X}_j(t)=
	\frac{\displaystyle\int_{\mathcal{I}_j(t)}
		x\rho_j(x,t)\,dx}
	{\displaystyle\int_{\mathcal{I}_j(t)}
		\rho_j(x,t)\,dx},
	\qquad j=1,2.
\end{equation}
The window has a fixed half-width of $1.2$ and follows the localized core throughout the evolution, reducing contributions from dispersive radiation.

We apply a nine-frame moving average (over an interval around 0.01T) of $\widehat{X}_j(t)$ to obtain $X_j(t)$, and calculate the separation as
\begin{equation}
D(t)=|X_2(t)-X_1(t)|,
\end{equation}
from which, we can also calculate the core velocities
\begin{equation}
	v_j(t)=\frac{dX_j(t)}{dt}
\end{equation}
using numerical differentiation.

\subsection{Localized interaction relative-motion energies}
\label{app:core_energy}

For the energy diagnostics in the intercomponent collisions, we introduce a separate window
with smoothly tapered edges, centered at each tracked core:
\begin{equation}
	w_j(x,t)=
	\begin{cases}
		1, & r_j\le h-\delta,\\
		\displaystyle\frac12
		\left[1+\cos\left(
		\frac{\pi[r_j-h+\delta]}{\delta}
		\right)\right],
		& h-\delta<r_j<h,\\
		0, & r_j\ge h,
	\end{cases}
\end{equation}
where $r_j=|x-X_j(t)|$, $h=1.6$, and $\delta=0.25$.
The corresponding localized norms are
\begin{equation}
	N_j^{\mathrm c}(t)=
	\int w_j(x,t)\rho_j(x,t)\,dx.
	\label{eq:core_norm}
\end{equation}
Although the total norm of each component is conserved,
$N_j^{\mathrm c}$ can vary as density enters or leaves
the moving window.
The localized intercomponent interaction energy is
\begin{equation}
	E_{12}^{\mathrm c}(t)=
	g_{12}\int w_{\cup}(x,t)
	\rho_1(x,t)\rho_2(x,t)\,dx,
	\qquad
	w_{\cup}=\max\{w_1,w_2\}.
\end{equation}
For attractive interaction, we write the magnitude of the
localized interaction energy as
$E_{\mathrm{bind}}(t)=-E_{12}^{\mathrm c}(t)$ and define
\begin{equation}
	\eta(t)=
	\frac{E_{\mathrm{bind}}(t)}
	{K_{\mathrm{rel}}(t)+\varepsilon_K},
	\qquad \varepsilon_K=10^{-14},
	\label{eq:capture_ratio_app}
\end{equation}
where $K_{\mathrm{rel}}$ is the relative kinetic-energy
estimate derived below, and $\varepsilon_K$ regularizes
the ratio as the relative velocity approaches zero.
For $K_{\mathrm{rel}}\gg\varepsilon_K$, $\eta$ compares
the magnitude of the attractive interaction energy with
the estimated energy of relative translation.
Large post-collision values of $\eta$ primarily reflect
the suppression of relative motion. Its fluctuations can
arise from core oscillations, deformation, and radiation,
and are amplified by the small denominator.
Numerical safeguards are included to avoid division by
vanishing denominators in normalized quantities.

To construct $K_{\mathrm{rel}}$, we approximate the localized
cores as rigid wave packets, neglecting internal currents
and time-dependent profile deformation. In the dimensionless
units used here, the kinetic-energy operator is
$-\frac12\partial_x^2$, and we adopt the ansatz
\begin{equation}
	\psi_j(x,t)=
	u_j\bigl(x-X_j(t)\bigr)
	e^{i v_j(t)[x-X_j(t)]+i\theta_j(t)},
	\qquad j=1,2,
\end{equation}
where $u_j$ is a real localized profile,
$v_j=\dot{X}_j$ is the core velocity, and $\theta_j(t)$
is a spatially uniform phase. The localized norm is
defined by Eq.~\eqref{eq:core_norm}.
Within this approximation, the gradient energy separates
into profile and translational contributions:
\begin{equation}
	E_{\mathrm{grad},j}
	=
	\frac12\int w_j(x,t)|\partial_x\psi_j|^2\,dx
	=
	\frac12\int w_j(x,t)|\partial_xu_j|^2\,dx
	+\frac12N_j^{\mathrm c}v_j^2.
\end{equation}
The translational contribution assigns each core a mass
$M_j=N_j^{\mathrm c}$ within the adopted ansatz, giving
\begin{equation}
	K_{\mathrm{tr}}
	=
	\frac12M_1v_1^2+\frac12M_2v_2^2.
\end{equation}
At each instant, we define the mass-weighted velocity
and the relative velocity as
\begin{equation}
	V=\frac{M_1v_1+M_2v_2}{M_1+M_2},
	\qquad
	v_{\mathrm{rel}}=v_2-v_1.
\end{equation}
The individual velocities are then
\begin{equation}
	v_1=V-\frac{M_2}{M_1+M_2}v_{\mathrm{rel}},
	\qquad
	v_2=V+\frac{M_1}{M_1+M_2}v_{\mathrm{rel}}.
\end{equation}
Substituting these expressions into $K_{\mathrm{tr}}$
separates the common and relative motion:
\begin{equation}
	K_{\mathrm{tr}}
	=
	\frac12(M_1+M_2)V^2
	+\frac12M_{\mathrm{red}}v_{\mathrm{rel}}^2,
\end{equation}
where
\begin{equation}
	M_{\mathrm{red}}(t)
	=
	\frac{N_1^{\mathrm c}(t)N_2^{\mathrm c}(t)}
	{N_1^{\mathrm c}(t)+N_2^{\mathrm c}(t)}
\end{equation}
is the reduced mass within this approximation.
We therefore use
\begin{equation}
	K_{\mathrm{rel}}(t)
	=
	\frac12M_{\mathrm{red}}(t)
	[v_2(t)-v_1(t)]^2
	\label{eq:relative_kinetic_estimate}
\end{equation}
as an estimate of the energy associated with relative
translation. It excludes the common motion and vanishes
when the two cores have equal velocities.

In the driven lattice, band dispersion, internal deformation,
and radiation can alter the relation between core velocity
and energy. Density redistribution within the finite tracking
windows can also affect the measured velocities.
Equation~\eqref{eq:relative_kinetic_estimate} is thus an
approximate diagnostic of relative motion, not a conserved
energy or a sufficient criterion for capture.

\subsection{Localized populations and overlap energy}
\label{app:coherent_diagnostics}

For vector-soliton collisions, we calculate the localized particle numbers
\begin{equation}
	\label{eq:coherent_local_norm}
	N_{j,n}(t)
	=
	\int \widetilde{W}_{n}(x,t)
	|\psi_j(x,t)|^2\,dx,
	\qquad j=1,2,\quad n=g,e,
\end{equation}
where $\widetilde{W}_{g,e}$ are moving rectangular windows following the tracked ground- and excited-state positions $X_g(t)$ and $X_e(t)$, with bounds $[X_g(t)-1,X_g(t)+1]$ and $[X_e(t)-1.5,X_e(t)+1]$, respectively. If the windows overlap, they are truncated at a common boundary to avoid double counting.
To identify the collision interval, we smooth the total density $\rho_1+\rho_2$ spatially using a Gaussian kernel with standard deviation $0.15$, obtaining $\bar{\rho}(x,t)$, and define
\begin{equation}
	R_{\mathrm{mid}}(t)
	=
	\frac{\bar{\rho}\!\left([X_g(t)+X_e(t)]/2,t\right)}
	{\min\!\left\{\rho_{g,\mathrm{pk}}(t),\rho_{e,\mathrm{pk}}(t)\right\}},
\end{equation}
where $\rho_{n,\mathrm{pk}}(t)=\max_{|x-X_n(t)|\le0.5}\bar{\rho}(x,t)$ is the local peak density near the tracked position. We exclude the entire interval from the first to the last sampled time satisfying $R_{\mathrm{mid}}(t)\ge0.30$, during which assigning overlapping density to the two wave packets becomes ambiguous. Outside this interval, the localized particle numbers are calculated from the original, unsmoothed densities. The post-collision populations provide a clearer characterization of the outcome.

To provide an intuitive understanding of the phase dependence, we define the overlap energy $E_{\mathrm{ol}}(D,\Delta\phi)$ of the two vector solitons at the collision time, with $D$ the manually imposed separation. 
Specifically, starting from the initial state
\begin{equation}
    {\tilde\Psi}_e(x)=\left(\begin{array}{c}
       \frac{\tilde\psi_{e}(x) }{\sqrt{2}} \\
       \frac{\tilde\psi_{e}(x) }{\sqrt{2}} 
    \end{array}\right), \quad     {\tilde\Psi}_g(x)=\left(\begin{array}{c}
       \frac{\tilde\psi_{g}(x) }{\sqrt{2}}e^{i\Delta\phi} \\
       \frac{\tilde\psi_{g}(x) }{\sqrt{2}} 
    \end{array}\right),
\end{equation}
we separately pump each vector soliton to the collision time $t_c$, which is determined by the coincidence of their centers. We denote the two instantaneous vector soliton as
\begin{equation}
    {\tilde\Psi}_e^c(x)=\left(\begin{array}{c}
       \frac{\tilde\psi_{e}^c(x) }{\sqrt{2}} \\
       \frac{\tilde\psi_{e}^c(x) }{\sqrt{2}} 
    \end{array}\right), \quad     {\tilde\Psi}_g^c(x)=\left(\begin{array}{c}
       \frac{\tilde\psi_{g}^c(x) }{\sqrt{2}}e^{i\Delta\phi} \\
       \frac{\tilde\psi_{g}^c(x) }{\sqrt{2}} 
    \end{array}\right). 
\end{equation}
We then evaluate the overlap energy
\begin{equation}
	\label{eq:coherent_overlap_energy}
	E_{\rm ol}(D,\Delta\phi)=E[{\tilde\Psi}_e^c(x)+{\tilde\Psi}_g^c(x-D)]-E[{\tilde\Psi}_e^c(x)]-E[{\tilde\Psi}_g^c(x-D)]
\end{equation}
where \(D\) is the manually imposed separation between the soliton centers. 
Here, \(E\) is the energy functional of the coupled system, evaluated at a fixed phase of the lattice drive. Explicitly, for \(\boldsymbol{\Psi}=(\Psi_1,\Psi_2)^{\mathsf T}\),
\begin{equation}
	\label{eq:coherent_energy_functional}
	\begin{aligned}
	E[{\Psi}]
	=
	\int dx\,\Bigg\{
	&\sum_{j=1}^{2}
	\left[
	\frac{1}{2}\left|\partial_x\psi_j\right|^2
	+V(x,t_c)|\psi_j|^2
	+\frac{g_{jj}}{2}|\psi_j|^4
	\right]
	+g_{12}|\psi_1|^2|\psi_2|^2
	\Bigg\},
	\end{aligned}
\end{equation}
Thus, \(E_{\mathrm{ol}}\) includes the overlap contributions from the kinetic, potential, and nonlinear interaction energies. 
The energy curves are obtained by static scans of the separation \(D\) and the phase \(\Delta\phi\). The lattice potential is held at \(V(x,t_c)\) throughout the scans. The resulting \(E_{\mathrm{ol}}\) characterizes the energetic effect of overlap for the prescribed profiles and provides an intuitive understanding of the phase dependence of the collisions.


	\end{appendix}
\end{widetext}

\bibliographystyle{apsrev4-2}
\bibliography{references}

\end{document}